\documentclass[10pt]{IEEEtran} 
\IEEEoverridecommandlockouts
\usepackage{multirow}
\usepackage{cite}
\usepackage{amsmath,amssymb,amsfonts}
\usepackage{algorithm}
\usepackage{algpseudocode}
\usepackage{textcomp}
\usepackage{caption}
\usepackage{graphicx}
\usepackage[table,xcdraw]{xcolor}
\usepackage{xcolor}
\def\BibTeX{{\rm B\kern-.05em{\sc i\kern-.025em b}\kern-.08em
    T\kern-.1667em\lower.7ex\hbox{E}\kern-.125emX}}
\begin{document}
\title{SSRESF: Sensitivity-aware Single-particle Radiation Effects Simulation Framework in SoC Platforms based on SVM Algorithm\\

}

\author{
\normalsize
\IEEEauthorblockN{Meng Liu\IEEEauthorrefmark{1}, Shuai Li\IEEEauthorrefmark{1}, Fei Xiao\IEEEauthorrefmark{1}, Ruijie Wang\IEEEauthorrefmark{1}, Chunxue Liu\IEEEauthorrefmark{2}, Liang Wang\IEEEauthorrefmark{2}}

\IEEEauthorblockA{\IEEEauthorrefmark{1}Faculty of Information Technology, School of Microelectronics, Beijing University of Technology, Beijing, China
}

\IEEEauthorblockA{\IEEEauthorrefmark{2}Beijing Microelectronics Technology Institute, Beijing, China}

Email: liumeng@bjut.edu.cn
\vspace{-1.1cm}
}

\maketitle

\begin{abstract}
    The ever-expanding scale of integrated circuits has brought about a significant rise in the design risks associated with radiation-resistant integrated circuit chips. Traditional single-particle experimental methods, with their iterative design approach, are increasingly ill-suited for the challenges posed by large-scale integrated circuits. In response, this article introduces a novel sensitivity-aware single-particle radiation effects simulation framework tailored for System-on-Chip platforms. Based on SVM algorithm we have implemented fast finding and classification of sensitive circuit nodes. Additionally, the methodology automates soft error analysis across the entire software stack. The study includes practical experiments focusing on RISC-V architecture, encompassing core components, buses, and memory systems. It culminates in the establishment of databases for Single Event Upsets (SEU) and Single Event Transients (SET), showcasing the practical efficacy of the proposed methodology in addressing radiation-induced challenges at the scale of contemporary integrated circuits. Experimental results have shown up to 12.78× speed-up on the basis of achieving 94.58\% accuracy. 
\end{abstract}


\section{Introduction}
By the integration degree is increasing, aerospace microprocessor as an important control component must have high radiation resistance. The research methods of single event effect (SEE) are including Carrying space experiment\cite{1} \cite{2}, Ground radiation experiment\cite{3} and Fault simulation experiment\cite{4}. The first two methods are limited by the factors of experimental funds, experimental cycles and cost, which makes it difficult to evaluate the radiation dependence capability of integrated circuits widely. Fault simulation technology can simulate the interaction of high-energy particles and chips at higher precision from the physical mechanism of single particle effect. Due to its performance ability and high speed, it is significant for assessing SEE of the circuit. However, with the development of VLSI technology, the design complexity has increased exponentially. A full-mode, single-particle experimental evaluation of a 1GB SDRAM device, for example, could occupy nearly 7.5 years. This considerable duration renders comprehensive evaluation and worst-case scenario identification virtually impossible. Moreover, existing fault simulation methods, heavily reliant on theoretical models\cite{5} and computational algorithms\cite{93}, often overlook the actual working environment of circuits and changes during operation. Furthermore, previous single-particle effect studies predominantly focus on individual functional modules such as RAM\cite{6}, CPU\cite{9}, and FPGA\cite{7}, leaving a knowledge gap in the context of System-on-Chip (SoC) designs. With the miniaturization trend and the growing prevalence of SoC designs, a holistic approach to single-particle effect analysis becomes increasingly necessary. SoCs, integrating diverse functional units into a single chip, present a complex environment where sensitivity to single-particle effects varies significantly among different modules.

In response to these challenges, we propose a sensitivity-aware single-particle radiation effect simulation framework (SSRESF) based on SVM algorithm for SoC platforms. This paper makes the following novel contributions as follows: 

\begin{itemize}
  \item The development of a single-particle radiation effect simulation framework for fault injection, simulation, and automatic soft error analysis for gate-level netlists.
  \item The introduction of a rapid circuit node classification method leveraging the SVM algorithm to expedite single-particle effect simulation.
  \item The implementation of dynamic simulation and data analysis for RISC-V PULP SoC, demonstrating up to 12.78× speed-up on the basis of an average accuracy of 94.58\%.
\end{itemize}

The paper unfolds as follows: Section II presents the background of single-particle radiation effect simulation. Section III details the proposed SSRESF. Section IV showcases the experimental results, and Section V concludes the paper.

\section{Background}
\subsection{Radiation Effect and Soft Errors}
Radiation effects on electronic systems can be divided into Total Ionizing Dose (TID), Displacement Damage (DD), and SEE. SEE refers to the impact of strongly ionizing particles on sensitive areas of integrated circuits, causing temporary disruptions in the circuit's standard operation, a critical factor in aerospace electronic systems. In integrated circuits, memory elements store data, with each bit representing a binary value. However, an SEE can trigger energy deposition and ionization phenomena when a single particle traverses a sensitive area of an integrated circuit, leading to bit flips in the memory element, a situation termed a Single Event Upset (SEU)\cite{10}. Furthermore, signals in a circuit traverse through different paths. A single particle crossing a sensitive area may cause a temporary alteration in charge distribution, producing a transient current pulse that propagates along the signal path, a phenomenon known as a Single Event Transient (SET)\cite{11}.
	In digital systems, bit flips caused by SEE are considered system-level effects or soft errors. While these bit flips are temporary and non-damaging to the hardware, they can disrupt the system's standard operation, leading to data or program execution errors.
	Thus, SEE events and associated phenomena such as bit flips and current pulses are considered system-level effects in digital systems, commonly referred to as soft errors. These errors significantly affect the design and operation of integrated circuits, particularly in high-reliability electronic systems like aerospace.

\subsection{Related Work}
The accelerating pace of research on radiation effect simulation with the growing complexity and performance demands of computing systems. Researchers are actively investigating and developing various acceleration methods to enhance the efficiency of soft error simulation and evaluation. The method introduced in \cite{12} expedites the process of transmission fault injection and propagation. This is achieved by concurrently processing multiple transmission fault injection and propagation tasks on the GPU, and parallel processing of gates at the same circuit level. While this approach significantly improves the efficiency of soft error estimation, it may increase complexity, energy consumption, and design difficulty compared to data analysis acceleration at the software level.  In \cite{13} , the authors propose a method for accelerating soft error analysis in large-scale circuits. This involves using different data structures for different hardware to minimize memory usage, and a two-stage heterogeneous simulation algorithm to further enhance performance. However, this method presents challenges including increased complexity, higher resource requirements, and difficulties in scalability and validation. In \cite{14} , machine learning is applied to aspects of fault annotation, particularly in virtualized fault annotation testing for safety-critical embedded systems. The focus is on the automated test case generation methodology, with machine learning primarily utilized in pre-fault annotation. Our proposed solution also focuses on single-particle soft error analysis and speeds up single-particle effect simulations by introducing a fast circuit node classification method using SVM algorithms.

\section{Proposed Methodology}
\subsection{Overview}
\begin{figure}[t]
    \centering
    \includegraphics[scale=0.55]{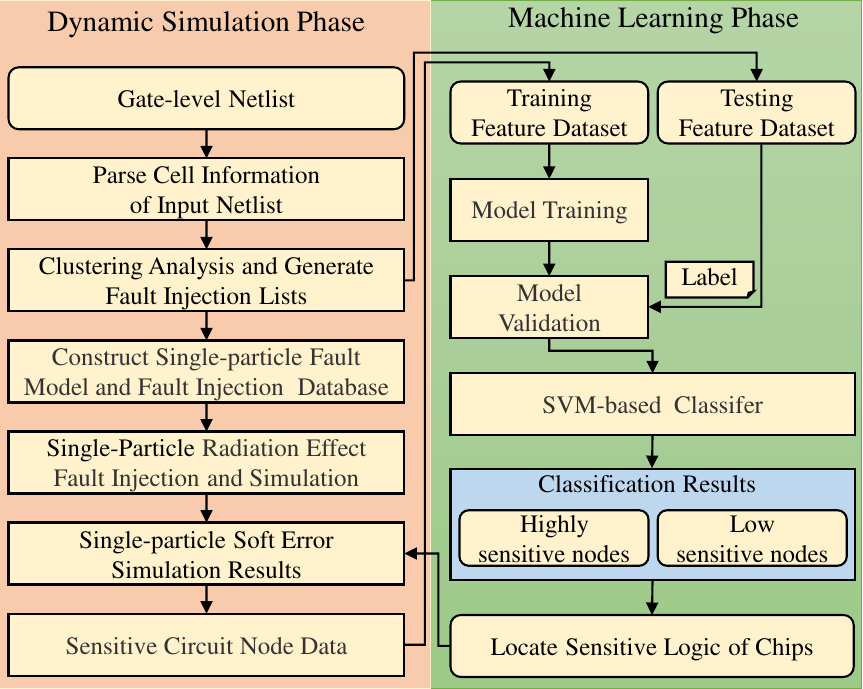}
    \caption{The overview of the SSRESF framework.}
    \label{overview}
\end{figure}

The overall flow of our SSRESF framework, as depicted in Fig.~\ref{overview}, is divided into two parts. The orange block diagram exhibits the analysis of the single-particle effect grounded in dynamic simulation, and the green block diagram depicts the classification of sensitive circuit nodes hinged on machine learning. Initially, an analysis method founded on a clustering algorithm is implemented for gate-level netlist circuits to extract the pertinent information of netlist units and categorize the circuits into multiple clusters. This approach enables the analysis of the functional relevance of the internal units of the circuit. Leveraging the single-particle flip-flop and transient characteristics of SoCs, we develop a fault injection model capable of simulating heavy-ion incidence with varying linear energy transfer (LET) values and fluxes. Concurrently, a SET and SEU database is established. Single-particle soft errors are automatically injected and simulated through linkage with the Verilog Procedural Interface (VPI) hardware interface. By integrating statistical analysis and fault injection simulation, the fault propagation characteristics within the SoC chip are simulated, enabling the acquisition of the SoC chip's sensitive node list data for the automatic analysis of single-particle soft errors. Given the results of radiation resistance effect simulation data, we apply feature engineering methods for data preprocessing design, tailored towards sensitive circuit nodes. This involves data cleaning, coding, normalization, among other processes. Through the application of the SVM algorithm, we conducted model training on the sensitive node list data, enabling us to achieve detailed classification prediction of the sensitive circuit nodes. This method empowers us to accurately locate specific areas within the SoC that are more susceptible to the impact of single-particle radiation.
\subsection{Clustering Analysis}
\begin{algorithm}
\small
\caption{Clustering Analysis for Internal Cells of Input Netlist}
\textbf{Input:} cellInfo (internal cells information), KN (number of clusters), LN (layer depth)\\
\textbf{Output:} clusters (Clusters of internal cells)
\begin{algorithmic}[1]

\Function{ClusteringAnalysis}{cellInfo, KN, LN}

\State $centers$ $\gets$ $random\_select(KN, cellInfo)$
\While{$centers$ change}
  \State $clusters$ $\gets$ $assign\_cells(cellInfo, centers, LN)$
  \State $centers$ $\gets$ $update\_centers(clusters, LN)$
  \EndWhile
\State \textbf{return} $clusters$
\EndFunction
\Function{assign\_cells}{CellInfo, centers, LN}
\State $clusters$ $\gets$ []
\For{each cell in $cellInfo$}
  \State $cluster$ $\gets$ $nearest\_center(centers, cell, LN)$
  \State Add cell to $cluster$
\EndFor
\State \textbf{return} $clusters$
\EndFunction
\Function{update\_centers}{clusters, LN}
\State $new\_centers$ $\gets$ []
\For{each $cluster$ in $clusters$}
  \State $center$ $\gets$ cell in $cluster$ with min sum of distance(cell, other cells in $cluster$, LN)
  \State Add $center$ to $new\_centers$
\EndFor
\State \textbf{return} $new\_centers$
\EndFunction
\end{algorithmic}
\end{algorithm}

Considering the intricate nature and vast mass of SoC circuits' internal resources, logical partitioning of these resources, is critical. We propose a clustering algorithm to group SoC internal cells. Our process involves cell information extraction from the circuit netlist. Analysis and clustering of functionally related internal chip cells, followed by fault injection simulation. We apply a cluster analysis algorithm that leverages the hierarchical structure and logical connections of the internal cells, enabling us to categorize cells based on their functional relevance. The objective is to assign functionally similar cells to the same category, allowing a more rational division of the chip's internal resources. Employing an equal-proportional random sampling strategy within these clusters, we optimize fault injection sample selection and distribution, thereby enhancing simulation efficiency.

In our clustering algorithm (see Algorithm 1), the inputs encompass cellInfo (information about the internal cells), KN (the number of clusters), and LN (the layer depth). The output is 'clusters', which are the results of the internal cells' clustering. Initially, clustering is conducted based on the hierarchical relationship of the cells. From the provided internal cell information, we randomly select a certain number of initial clustering centers. Utilizing the acquired cell information, we define the inter-cluster distance function as follows:
\begin{equation}
\small
D_{A,B}=\sum_{Li=1}^{LN} Compare(Module_{ALi},Module_{BLi})*2^{LN-Li}
\end{equation}
Here, $D_{A,B}$ is the distance function between two cells A and B, LN is the layer depth, and Li denotes the i-th layer where cells A, B are located. $Module_{ALi}$ and $Module_{BLi}$ represent the layer module to which cell A and B belong when they are in the lower Li layer. The $Compare(Module_{ALi},Module_{BLi})$ function is used to examine whether $Module_{ALi}$ and $Module_{BLi}$ are the same. If they are identical, it returns 0; otherwise, it returns 1. The initial clustering result is achieved by calculating the distance between each cell and the cluster center, and categorizing it into the cluster with the shortest distance (lines 9-16 in Algorithm 1). Then, during the iteration process, we recalculate the sum of the distances of the cells within each cluster based on the current clustering result. The cell with the smallest sum of distances is selected as the new cluster center (lines 17-24 in Algorithm 1). Subsequently, we perform reallocation and clustering of the cells again, based on the new cluster centers. This iterative process continues until the cluster centers no longer change. Ultimately, we obtain the clustering results of the internal cells (lines 1-8 in Algorithm 1). The clustering result is the division of the cells into multiple clustered clusters. The cells within each cluster are more similar to each other, and the cells between different clusters are less similar. This clustering result aids in the analysis of the functional relevance of the cells within the circuit.

\subsection{Fault Model}
\begin{figure}[th]
    \centering
    \includegraphics[scale=0.85]{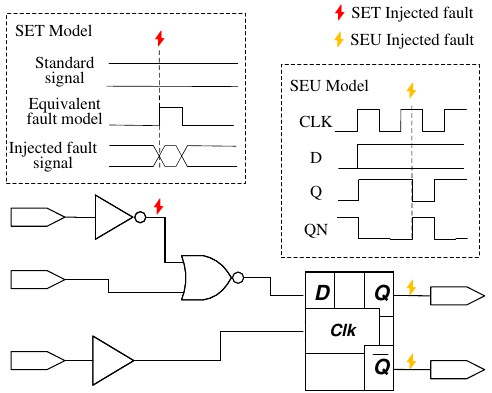}
    \caption{The SET and SEU single-particle fault models.}
        \label{pic2.png}
\end{figure}
\begin{figure}[th]
    \centering
    \includegraphics[scale=0.5]{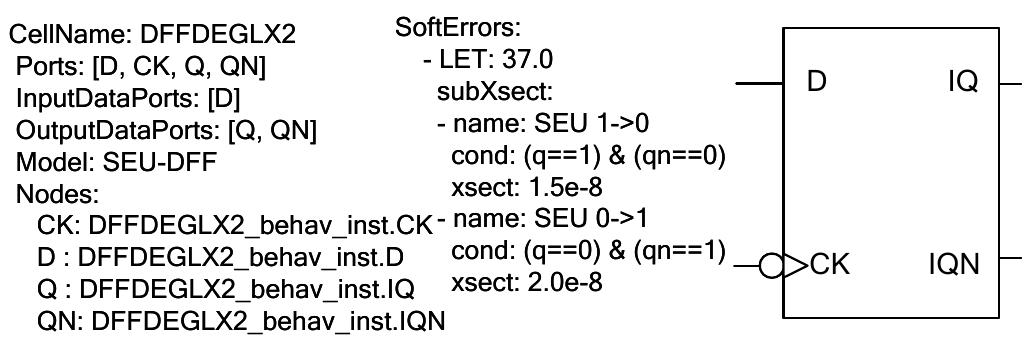}
    \caption{The SET and SEU single-particle soft error database.}
    \label{pic3.png}
\end{figure}

Upon completing the cluster analysis, we generate a corresponding fault injection list for each cluster. For every cluster, we construct an appropriate fault model based on the types of internal units it contains for fault injection simulation.
The construction method for the single-particle equivalent model of each unit is depicted in Fig. \ref{pic2.png}. The red segment of Fig. \ref{pic2.png} represents the modeling method for the SET effect of the combinational logic unit. This entails constructing and injecting the model by developing an equivalent square-wave model to be injected into the target line network. In contrast, the yellow segment of Fig. \ref{pic2.png} showcases the modeling method for the SEU effect of the rising edge triggered timing logic unit. After fault injection, the signal produces 0→1 and 1→0 value transitions. The duration of the flip-flop effect of the timing logic is associated with both the random moment when the fault injection occurs and the moment when the driving clock edge arrives.
Based on the construction of the single-particle fault model, we designed a SET and SEU single-particle soft error database, as illustrated in Fig. \ref{pic3.png}. This database contains the node information and soft error injection information of a D-Flip flops cell.
For each cell, we define the flip-flop cross-section values for soft error injection based on various LET values. We selected LET values of 1.0, 37.0, and 100.0 to encompass different radiation environments. These flip-flop cross-section values, which are based on experimental data and simulation results, are designed to simulate the soft-error susceptibility of various circuit units induced by single-particle radiation.
\subsection{Fault Injection and Simulation}
Our fault injection and simulation methodology is founded on the use of generic simulators. Based on the fault injection list, we apply the constructed fault model and simulate the sampling unit for fault injection using Synopsys VCS and OSS CVC Verilog simulators. Interaction with the simulators is managed through the generic IEEE Std 1364-2005 standard (Verilog) and the VPI invocation of the single-particle error database. These tools enable us to interact with and control different types of standard-compliant simulators.

For each specific cluster and its corresponding fault injection list derived from our cluster analysis, we perform fault injection and simulation sequentially for the circuit units. The fault injection location is localized at the output port of each internal unit. We select random time points for fault injection and continuously monitor the chip's main output signals by comparing the VCD files generated from the post-fault injection simulation. When an abnormal output signal is detected following fault injection, we classify it as a soft error. If no abnormality is observed, we proceed with the fault injection simulation for the next cell in the list. Upon completing the fault injection simulation for all cells in the list, we tally the number of soft errors for each cluster. During the soft error rate (SER) evaluation stage, we assess the SER of the entire cluster based on the sampled units therein. For each cluster, we calculate the SER of its sampling unit, determined by the ratio of the number of observed soft errors in the fault injection simulation to the total number of sampled cells. We then consider the SER of the sampling units as the SER for that cluster. After multiple iterations of soft error simulation and analysis for different clusters, we obtain the evaluation results of SER for all clusters. Ultimately, this provides us with the SER of the entire SoC. The formula for the SER is given as:
\begin{equation}
\small
SER_{Chip}=\frac{ {\textstyle \sum_{i=0}^{KN}}Cell_{Cluster_i} \times SER_{Cluster_i}}{\textstyle \sum_{i=0}^{KN}CellN_{Cluster_i}} 
\end{equation}
Based on the assessment of the SER of each cluster, we sort the clusters in ascending order of soft error probability. The high probability clusters are highly sensitive to single-particle effects, and the circuit nodes of their internal units constitute the sensitive circuit node list data.
\subsection{Classification of Sensitive Circuit Nodes}
When facing the problem of single-particle sensitivity analysis of large-scale circuits, we need to accelerate the work of sensitive node classification because the traditional traversal simulation methods incur high computational and time overheads. 
The SVM algorithm emerges as a powerful tool for this classification task because it seeks a hyperplane that maximizes the margins between different classes. In our context, this means that SVM can effectively distinguish between sensitive and non-sensitive circuit nodes based on the training data.
\begin{figure}[t]
    \centering
    \includegraphics[scale=0.6]{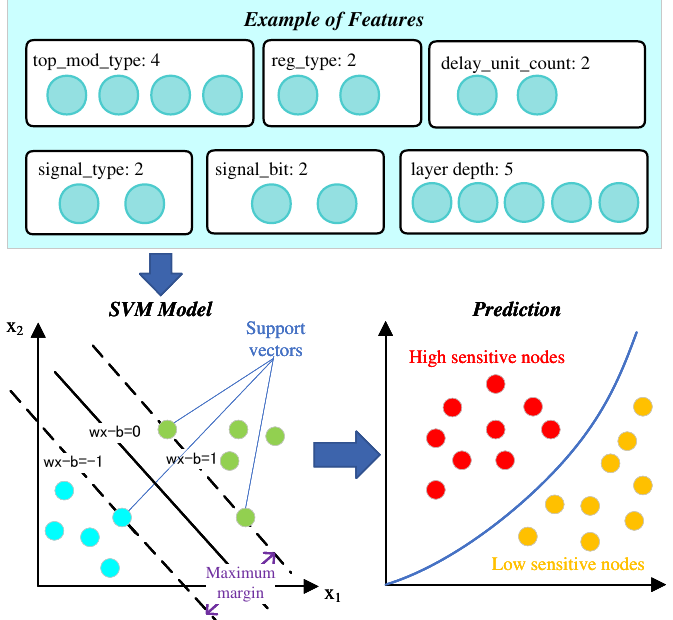}
    \caption{The classification process for sensitive circuit nodes.}
    \label{pic4.png}
\end{figure}
The overall process is depicted in Fig. \ref{pic4.png}, where we obtain the structural features of the netlist via feature extraction methods. We then feed this dataset into the SVM model for training, and eventually, we obtain the sensitivity prediction results for the circuit nodes.

Our first step is to analyze the obtained sensitive circuit node list data by frequency count to understand the distribution pattern of circuit nodes with different sensitivity degrees. After cleaning, coding, and normalizing the data through feature engineering methods, we construct the feature database. We then define the classification rules for the sensitivity degree of circuit nodes, manually classify the data, and use these classification results as the labels for circuit nodes. We apply the SVM classifier to learn from the training dataset. The circuit node data from the fault injection list forms the test dataset, which we use to validate the learned model and obtain the final model. This adequately trained prediction model can then be used externally to provide a sensitive data prediction service. This service is capable of distinguishing circuit nodes into high sensitivity and low sensitivity categories, enabling fast classification of sensitive circuit nodes.

\section{Experiment Results}

\begin{table*}[h]
\renewcommand\arraystretch{1.2}
\caption{Soft error results for different functional modules of benchmark.}
\label{tab1}
\centering
\resizebox{\textwidth}{!}{
\begin{tabular}{|c|ccc|ccc|ccc|ccc|}
\hline
\multirow{2}{*}{\textbf{Benchmark}} & \multicolumn{3}{c|}{\textbf{Memory Info}}                                                                                                                                                            & \multicolumn{3}{c|}{\textbf{Bus Info}}                                                                                                                                                           & \multicolumn{3}{c|}{\textbf{CPU Logic Info}}                                                                                                                                                           & \multicolumn{3}{c|}{\textbf{Xsect Info}}                                                                                                                                                                        \\ \cline{2-13} 
                                    & \textbf{\begin{tabular}[c]{@{}c@{}}Memory\\ Type\end{tabular}} & \textbf{\begin{tabular}[c]{@{}c@{}}Memory\\ Size\end{tabular}} & \textbf{\begin{tabular}[c]{@{}c@{}}Memory\\ SER (\%)\end{tabular}} & \textbf{\begin{tabular}[c]{@{}c@{}}Bus\\ Type\end{tabular}} & \textbf{\begin{tabular}[c]{@{}c@{}}Bus bit\\ Width\end{tabular}} & \textbf{\begin{tabular}[c]{@{}c@{}}Bus\\ SER (\%)\end{tabular}} & \textbf{\begin{tabular}[c]{@{}c@{}}CPU\\ Type\end{tabular}} & \textbf{\begin{tabular}[c]{@{}c@{}}Number of\\ CPU Cores\end{tabular}} & \textbf{\begin{tabular}[c]{@{}c@{}}CPU\\ SER (\%)\end{tabular}} & \textbf{\begin{tabular}[c]{@{}c@{}}Number\\ of clusters\end{tabular}} & \textbf{\begin{tabular}[c]{@{}c@{}}SET\\ Xsect (cm²)\end{tabular}} & \textbf{\begin{tabular}[c]{@{}c@{}}SEU\\ Xsect (cm²)\end{tabular}} \\ \hline
PULP SoC\_1                         & SRAM                                                           & 64KB                                                           & 0.09                                                               & APB                                                         & 8                                                                & 0.11                                                            & RV32I                                                       & 1                                                                      & 0.03                                                            & 5                                                                     & 1.10E-03                                                           & 1.30E-03                                                           \\
PULP SoC\_2                         & DRAM                                                           & 64KB                                                           & 0.05                                                               & APB                                                         & 16                                                               & 0.19                                                            & RV32I                                                       & 2                                                                      & 0.05                                                            & 6                                                                     & 2.30E-03                                                           & 2.70E-03                                                           \\
PULP SoC\_3                         & SRAM                                                           & 256KB                                                          & 0.35                                                               & AHB                                                         & 32                                                               & 0.41                                                            & RV32IM                                                      & 1                                                                      & 0.07                                                            & 8                                                                     & 3.20E-03                                                           & 4.30E-03                                                           \\
PULP SoC\_4                         & DRAM                                                           & 256KB                                                          & 0.2                                                                & AHB                                                         & 64                                                               & 0.47                                                            & RV32IM                                                      & 2                                                                      & 0.15                                                            & 9                                                                     & 4.50E-03                                                           & 5.50E-03                                                           \\
PULP SoC\_5                         & SRAM                                                           & 1MB                                                            & 0.43                                                               & AXI                                                         & 128                                                              & 0.61                                                            & RV32IMF                                                     & 1                                                                      & 0.11                                                            & 14                                                                    & 5.40E-03                                                           & 7.10E-03                                                           \\
PULP SoC\_6                         & DRAM                                                           & 1MB                                                            & 0.35                                                               & AXI                                                         & 256                                                              & 0.73                                                            & RV32IMF                                                     & 2                                                                      & 0.2                                                             & 15                                                                    & 6.50E-03                                                           & 8.30E-03                                                           \\
PULP SoC\_7                         & SRAM                                                           & 2MB                                                            & 0.75                                                               & APB                                                         & 512                                                              & 0.87                                                            & RV32IMAFD                                                   & 1                                                                      & 0.15                                                            & 18                                                                    & 7.60E-03                                                           & 9.90E-03                                                           \\
PULP SoC\_8                         & DRAM                                                           & 2MB                                                            & 0.65                                                               & APB                                                         & 1024                                                             & 1.05                                                            & RV32IMAFD                                                   & 2                                                                      & 0.23                                                            & 19                                                                    & 8.90E-03                                                           & 1.11E-02                                                           \\
PULP SoC\_9                         & SRAM                                                           & 4MB                                                            & 1.05                                                               & AHB                                                         & 2048                                                             & 1.19                                                            & RV64I                                                       & 1                                                                      & 0.19                                                            & 21                                                                    & 9.80E-03                                                           & 1.27E-02                                                           \\
PULP SoC\_10                        & Rad-hard SRAM                                                  & 4MB                                                            & 0.03                                                               & AHB                                                         & 4096                                                             & 1.39                                                            & RV64I                                                       & 2                                                                      & 0.25                                                            & 23                                                                    & 1.12E-02                                                           & 1.38E-02                                                           \\ \hline
\end{tabular}
}
\end{table*}

The experiments were performed on a 64-bit Linux machine with a 24-core Intel Xeon CPU (2.20 GHz), 64 GB of RAM, and an Nvidia RTX 3090 Ti GPU. To validate the accuracy of our single-particle effect simulations, we employed the commercial simulator Synopsys VCS 2019 as well as the open-source simulator OSS-CVC. These tools allowed us to confirm the reliability of our results. We considered using the Verilator and Icarus Verilog for simulation. However, Verilator was not suitable because it does not support gate-level netlist simulation. Icarus Verilog, although a viable option, was eventually ruled out due to its slower simulation speed.


\subsection{Soft Error Analysis}
We selected for the experiments has been the RISC-V PULP SoC designed by Swiss Federal Institute of Technology in Zurich\cite{15}, since it was already successfully integrated into other platforms. We conducted cluster analysis, fault injection, and simulation experiments on gate-level netlists for 10 different compositions of RISC-V PULP SoC. We tabulated the the SER for different functional modules of these netlists, and the SET and SEU soft error intercepts for each netlist are counted in Table \ref{tab1}. The objective of these experiments was to assess the single-particle sensitivities of different functional modules in each SoC using our method. This assessment helps to identify the most sensitive modules.  
The experimental results indicate that the SER of the bus is typically the highest among all SoCs, followed by memory, and finally CPU logic. From a memory perspective, SRAM is more prone to single-particle flip-flops than DRAM. This is because SRAM's individual memory cell consists of only a few transistors, while DRAM's individual memory cell has a capacitive structure, operates at a higher voltage, and is less susceptible to direct single-particle flip-flops. As the memory size increases, so does the soft error probability, especially with DRAM. However, SoC\_10, which uses Radiation-hardened (Rad-hard) SRAM as memory, shows a lower SER, indicating that Rad-hard SRAM is more resistant to single-particle effects. Additionally, we discovered that the SER of the bus increases as the bus bit width increases. In multi-core CPUs, there is a higher SER due to interactions and increased complexity. In conclusion, as the SoC's complexity increases, so does the number of clusters and the soft-error cutoff. This suggests that more complex SoCs are more susceptible to single-particle-induced soft errors.

\subsection{Classification of Sensitive Circuit Nodes}
\begin{figure}[b]
    \centering
    \includegraphics[scale=0.72]{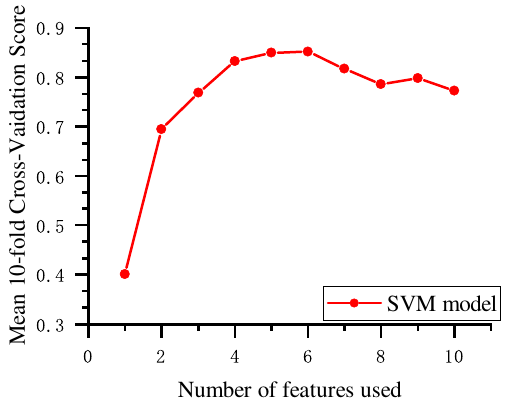}
    \caption{The features selection of SVM model.}
    \label{pic5.png}
\end{figure}

\begin{table}[h]
\caption{Results of SVM classification.\label{tab2}}
\centering
\renewcommand\arraystretch{1.1}
\resizebox{\linewidth}{!}{
\begin{tabular}{cccccc}
\hline
\textbf{Benchmark} & \textbf{TNR} & \textbf{TPR} & \multicolumn{1}{l}{\textbf{Precision}} & \textbf{Accuracy} & \textbf{F1 Score} \\ \hline
PULP SoC\_1             & 90.43\%      & 83.56\%              & 87.14\%                                & 87.43\%      & 0.85              \\
PULP SoC\_2             & 91.44\%      & 83.67\%              & 88.49\%                                & 88.02\%      & 0.86              \\
PULP SoC\_3             & 89.84\%      & 81.51\%              & 86.23\%                                & 86.19\%      & 0.84              \\
PULP SoC\_4             & 91.98\%      & 85.03\%              & 89.29\%                                & 88.92\%      & 0.87              \\
PULP SoC\_5             & 91.44\%      & 83.67\%              & 88.49\%                                & 88.02\%      & 0.86              \\
PULP SoC\_6             & 90.91\%      & 82.31\%              & 87.68\%                                & 87.13\%      & 0.85              \\
PULP SoC\_7             & 90.37\%      & 84.35\%              & 87.32\%                                & 87.72\%      & 0.86              \\
PULP SoC\_8             & 91.49\%      & 82.19\%              & 88.24\%                                & 87.43\%      & 0.85              \\
PULP SoC\_9             & 90.43\%      & 84.25\%              & 87.23\%                                & 87.72\%      & 0.86              \\
PULP SoC\_10            & 90.96\%      & 82.88\%              & 87.68\%                                & 87.43\%      & 0.85              \\ \hline
Average               & 90.91\%      & 83.56\%              & 87.77\%                                & 87.69\%      & 0.86              \\ \hline
\end{tabular}
}
\end{table}
In this section, we apply the SVM algorithm based sensitive circuit node classification method to the 10 gate-level netlists of the SoC benchmarks noted in Table \ref{tab1}. The Python machine learning library, scikit-learn, is used to train our SVM classifier model. The first experiment performed was for feature selection. 
Fig. \ref{pic5.png} shows the relationship between the mean 10-fold cross-validation score and the number of features included in the model. It was observed that the mean 10-fold cross validation score reached the highest value when the number of features was 6. Therefore, this optimized subset of 6 features was used for all subsequent experiments. After feature selection, grid search was applied to optimize the hyper-parameters of the classifiers. In this case, a 10-fold cross-validation was also used and optimal pairs $(C,\gamma)$ were determined for SVM classifier. Next, we evaluated the performance of SVM classifiers by performing 10-fold cross validation. 
\begin{figure}[b]
    \centering
    \includegraphics[scale=0.25]{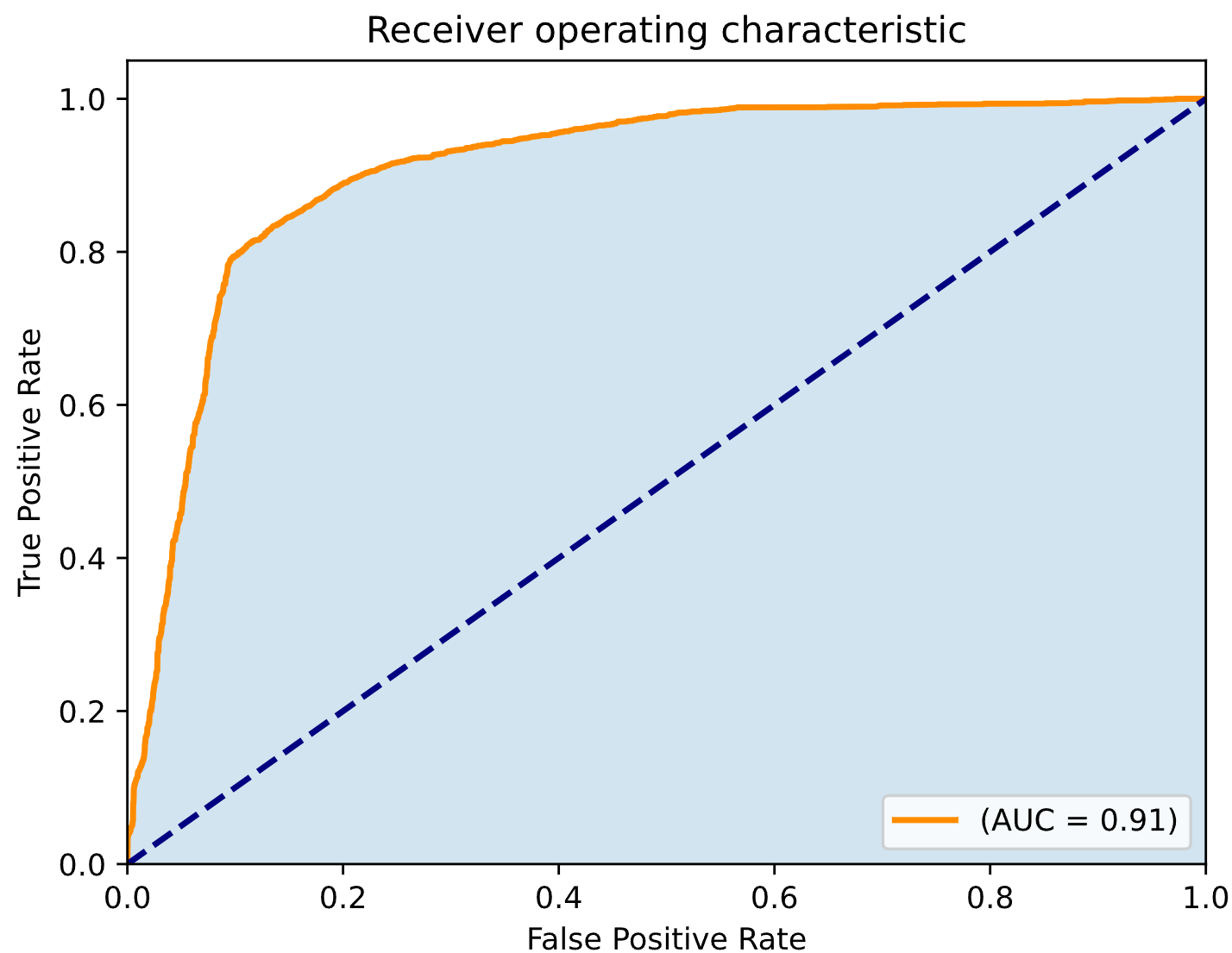}
    \caption{The ROC curve of SVM model.}
    \label{pic6.png}
\end{figure}
For binary classification, we evaluate the results using four key metrics: true positives (TP), true negatives (TN), false positives (FP), and false negatives (FN). These metrics facilitate the calculation of several performance indicators including true positive rate (TPR), true negative rate (TNR), precision, accuracy, and the F1 score. We utilize these metrics to assess the effectiveness of our method. Table \ref{tab2} displays the results of our experiment. It reveals that our method ensures a low false positive rate, with an average TNR of 90.91\% and a TPR of 83.56\%. Additionally, both precision and accurary reach 87\%. Fig. \ref{pic6.png} illustrate the Receiver Operating Characteristic (ROC) curve of our model. The closer the ROC curve aligns to the upper left corner, the better our model performs. Our experimental results underscore the effective application of machine-learning-based classification methods in classifying sensitive circuit nodes for gate-level netlists.

\begin{table}[t]
\caption{Runtime comparison among VCS, CVC and our model.\label{tab3}}
\centering
\renewcommand\arraystretch{1.2}
\resizebox{\linewidth}{!}{
\begin{tabular}{ccccccc}
\hline
\textbf{Flux} & \textbf{\begin{tabular}[c]{@{}c@{}}VCS \\ Simulation\\ Runtime(s)\end{tabular}} & \textbf{\begin{tabular}[c]{@{}c@{}}CVC \\ Simulation\\ Runtime(s)\end{tabular}} & {\color[HTML]{333333} \textbf{\begin{tabular}[c]{@{}c@{}}Model \\ Prediction\\ Time(s)\end{tabular}}} & {\color[HTML]{333333} \textbf{\begin{tabular}[c]{@{}c@{}}Speedup\\ (VCS)\end{tabular}}} & {\color[HTML]{333333} \textbf{\begin{tabular}[c]{@{}c@{}}Speedup\\ (CVC)\end{tabular}}} & \textbf{\begin{tabular}[c]{@{}c@{}}Model\\ Accuracy\end{tabular}} \\ \hline
4e8           & 170.24                                                                          & 200.23                                                                          & 23.95                                                                                                 & 7.11×                                                                                   & 8.36×                                                                                    & 80\%                                                              \\
5e8           & 200.22                                                                          & 280.25                                                                          & 24.08                                                                                                 & 8.31×                                                                                   & 11.64×                                                                                   & 100\%                                                             \\
6e8           & 310.33                                                                          & 300.30                                                                          & 24.19                                                                                                 & 12.83×                                                                                  & 12.41×                                                                                   & 100\%                                                             \\
7e8           & 300.31                                                                          & 330.29                                                                          & 23.82                                                                                                 & 12.61×                                                                                  & 13.87×                                                                                   & 100\%                                                             \\
8e8           & 380.38                                                                          & 410.41                                                                          & 23.26                                                                                                 & 16.35×                                                                                  & 17.64×                                                                                   & 92.9\%                                                            \\ \hline
Avg.          & 272.30                                                                          & 304.30                                                                          & 23.86                                                                                                 & 11.44×                                                                                  & 12.78×                                                                                  & 94.58\%                                                           \\ \hline
\end{tabular}
}
\end{table}

\subsection{Comparison}

\begin{figure}[th]
    \centering
    \includegraphics[scale=0.41]{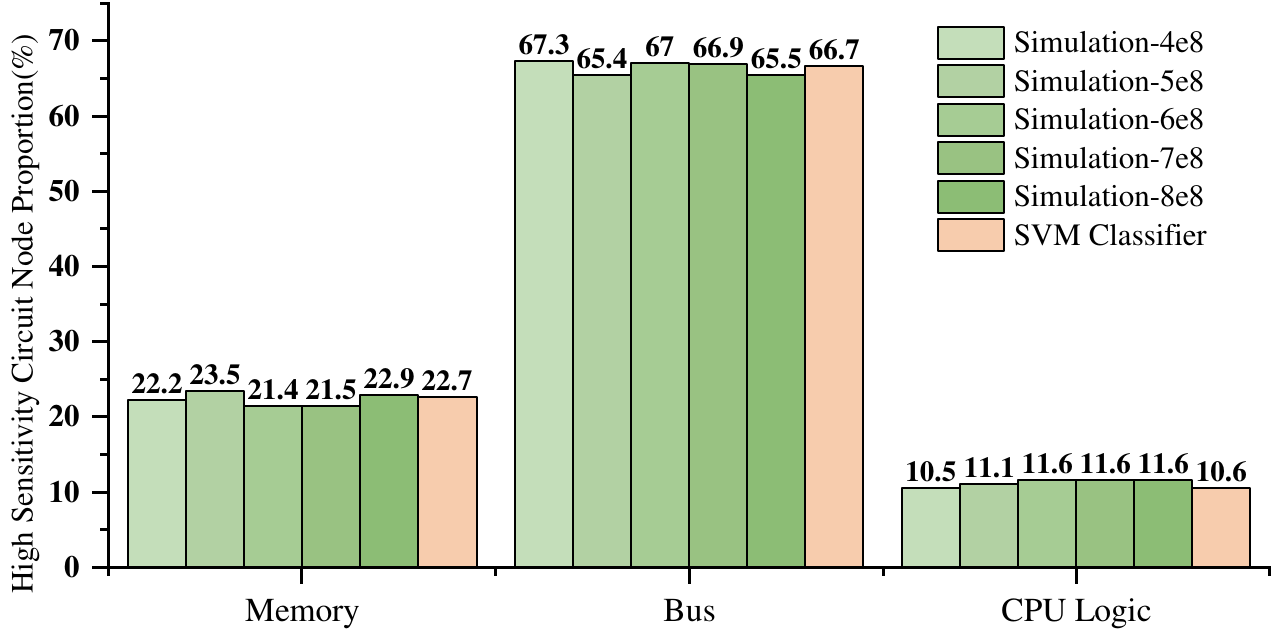}
    \caption{The proportion of high sensitivity circuit node.}
    \label{pic7.png}
\end{figure}

In this section, we contrast two methodologies using PULP SoC\_1 as a case study. We execute fault injection simulations under various flux conditions to acquire a subset of sensitive circuit nodes. Subsequently, SVM classifier is employed to target circuit nodes with unknown sensitivity in the netlist. We then count the number of highly sensitive nodes and measure the runtime of both methods. The distribution of highly sensitive nodes in different modules—bus, memory, and CPU logic—as predicted by the SVM classifier is illustrated in Fig. \ref{pic7.png}. We find that the distributions are highly consistent, i.e., the number of highly sensitive nodes in bus, memory, and CPU logic decreases sequentially. This is consistent with the results of the soft error analysis experiments. Table \ref{tab3} lists the combined comparison of fault injection and simulation using VCS and CVC with our SVM classifier in terms of runtime. The experimental results show that the average speedup ratio of the SVM classifier's runtime is 11.44 and 12.78 based on an average accuracy of 94.58\% compared to the VCS and CVC simulation results, respectively. This demonstrates the feasibility of SVM-based assisted single-particle effect simulation speedup.

\section{Conclusion}
In this paper, we introduce the SSRESF for sensitivity-aware single-particle radiation effects based on SVM algorithm. We propose a clustering algorithm to group the gate-level netlist based on functional similarity, enabling the extraction of cell information. The framework integrates SET and SEU databases, supporting single-particle effect fault injection, soft error analysis, and sensitive circuit node classification.  We conduct soft error analysis experiments on SoC using Synopsys VCS and OSS-CVC, and subsequently perform classification prediction experiments for sensitive nodes based on the soft error simulation results. This illustrates the variance in single-particle effect sensitivities among different modules in the SoC architecture. The results demonstrate that our method can accelerate single-particle effect simulation up to 12.78×, while maintaining an average accuracy of 94.58\%.

\bibliographystyle{IEEEtran}
\bibliography{ref}
\end{document}